\documentclass[letterpaper,12pt]{article}

\usepackage{endnotes}
\usepackage{amssymb}
\usepackage{amsfonts}
\usepackage{amsmath} 
\usepackage{amsthm}
\usepackage{graphicx,color}
\usepackage{graphicx, graphics, color, epsfig, setspace, epstopdf}
\usepackage{comment}
\usepackage{chicago}
\usepackage{array}

\usepackage{url} 

\usepackage{setspace}
\def\R{\mathbb{R}}
\def\endproof{\hfill\diamondsuit}

\def\sL{{\mathcal L}}

\def\tv{\tilde{v}}

\def\E{\mathbb{E}}
\def\V{\mathbb{V}}

\def\N{\mathbb N}

\numberwithin{equation}{section}
\theoremstyle{plain}                % title and  number in bold, text italic
\newtheorem{theorem}{Theorem}[section]
\newtheorem{lemma}[theorem]{Lemma}

\newtheorem{corollary}[theorem]{Corollary}
\theoremstyle{definition}           % title and number in bold, text normal
\newtheorem{definition}[theorem]{Definition}
\newtheorem{example}[theorem]{Example}

\theoremstyle{remark}               % title and number in italic, text normal

\usepackage{geometry}
\newgeometry{left=1.25in,right=1.25in,top=1.5in,bottom=1.5in}

\usepackage{ulem} 

\begin{document}

\pagestyle{empty}

\begin{center}
\large{\bf Is Kyle's equilibrium model stable?}$^\ast$
\makeatletter{\renewcommand*{\@makefnmark}{}\footnotetext{\hspace{-.35in} $^\ast${The authors have benefited from helpful comments from Kerry Back, Dan Bernhardt, Bradyn Breon-Drish, Vincent van Kervel, S. Vishwanathan., and participants at the SIAM meeting in Philadelphia  (2023). The corresponding author is Kasper Larsen. Kasper Larsen has email: KL756@math.rutgers.edu, and Umut \c{C}et\.in has email: U.Cetin@lse.ac.uk. }
\makeatother}}
\end{center}

\ \\

\begin{center}

{ \bf Umut \c{C}et\.in }\\ 

Department of Statistics 

London School of Economics

\ \\ \ \\

{ \bf Kasper Larsen}\\
Department of Mathematics

Rutgers University

%110 Frelinghuysen Road

%Piscataway, NJ 08854-8019, USA

%Phone: +1 848-445-2390
\ \\
{\normalsize \today }
\end{center}

\ \\

\begin{verse}
{\sc Abstract}: In the  dynamic discrete-time trading setting of Kyle (1985),  we prove that Kyle's equilibrium model is stable when there are one or two trading times. For three or more trading times, we prove that Kyle's equilibrium is not stable. These theoretical results are proven to hold irrespectively of all Kyle's  input parameters. 
\end{verse}

%\begin{verse}
%{\sc JEL codes}: G11, G12\ \\
%\end{verse}
\vspace{0.5cm}
\begin{verse}
\end{verse}

\newpage
\pagestyle{plain}
\addtocounter{page}{-1}

\section{Introduction} Kyle (1985) is a cornerstone model in today's market microstructure theory. Its  relevance is long established; see, e.g., the textbook discussions in Back (2017). We consider the  discrete-time formulation where an informed trader, noise traders,  and market makers dynamically trade the stock at $N\in\N$ time points.  After observing the aggregate order flow, the market makers set the stock price to clear the stock market.  Kyle (1985) proves existence of a unique linear equilibrium, and we study its stability properties. We prove that the number of trading time points $N\in\N$ determines all stability properties of Kyle's equilibrium. Specifically, irrespectively of all other input parameters, we prove that Kyle's equilibrium is stable for $N\in \{1,2\}$ and not stable for $N\ge3$.

Hadamard (1902) deems a model well-posed if existence, uniqueness, and stability hold. Kyle (1985) gives existence and uniqueness of a linear equilibrium, and we use the convergence of policy iterations to determine if Kyle's linear equilibrium is stable.\footnote{Using policy iterations to iteratively calculate optimizers is well established and is intimately related to the Bellman equation in optimal control theory, see, e.g., Theorem 3 in Chapter I.11 in Bellman (1957).}   We start the policy iterations from a marginal perturbation away from the insider's equilibrium orders. Then, we iteratively create a sequence of insider orders by considering  the market makers' response to the insider's perturbed orders and the subsequent response by the insider and so on. Kyle's equilibrium is deemed stable if this iteratively constructed sequence of insider orders converges to Kyle's equilibrium orders whenever the initial orders are only marginally different from Kyle's equilibrium orders. % In contrast, if no initial marginal order  perturbation away from the insider's equilibrium orders can produce a sequence of iterative insider orders that converge to Kyle's equilibrium orders, the equilibrium is deemed unstable.   
 This definition of  fixed-point stability in terms of iterations of a marginal perturbation away from  the fixed point itself is standard in numerical analysis, see, e.g., Definition 1.3 in the textbook S\"uli and Mayers (2003).\footnote{There are several related notions of stability. For example,  stability of the fixed-point operator itself is defined in Definition 7.1 in Berinde (2007), and stability for dynamical systems (such as ODE solutions) is defined in Definition 6.1 in Betounes (2010). Stability of  games is defined in Kohlberg and Mertens (1986). Robustness of games is defined Stauber (2006).} 

Defining stability in terms of the convergence of policy iterations is natural in the context of  a financial market equilibrium because policy iterations can be viewed as the best responses of rational agents given the current state of the market. Thus, a stable equilibrium  has the property that if the agents find themselves in the equilibrium's vicinity, their actions draw the economy closer to equilibrium\footnote{This notion of stability can be found in financial economics literature.  In DeMarzo, Kaniel, and Kremer (2008),  a stable equilibrium leads to a price bubble, which means that small shocks to the agents' beliefs may  result in departures from optimal risk sharing associated with typically non-stable equilibria.  Biais, Foucault,  and Moinas (2015)  focus on stable equilibria in their study of firms investing in fast trading technologies.}.  

To the best of our knowledge stability of Kyle's dynamic equilibrium model has not been studied in the literature. The closest study to our paper is Boulatov and Bernhardt (2015), who proves a robustness property for Kyle's equilibrium when $N=1$. We study stability of Kyle's discrete-time model in full generality.    In addition to being of theoretical interest, understanding stability properties of Kyle's model is relevant because algorithmic market makers have become an important part of asset pricing; see, e.g., Colliard, Foucault, and Lovo (2022).

We prove two theoretical results. First, when  $N\in \{1,2\}$, we prove that Kyle's equilibrium is always stable. Our proof establishes that the policy iterations are locally contracting near Kyle's equilibrium. Even stronger, for $N=1$, we show that Kyle's equilibrium is a super-attractive fixed point in the sense that local convergence is strictly faster than linear. Second, when $N\ge3$, we prove that Kyle's equilibrium is always not stable.   Table \ref{tab0} summarizes our theoretical results:\ \\

\vspace{-0.8cm}
\begin{table}[ht]
\centering  \caption{Key results}
\label{tab0}
\vspace{.2cm}
\begin{tabular}{c|c|c}
\hline\hline
$N$  & Fixed-point type& Conclusion\\
\hline\hline
1&Super-attractive& stable\\
\hline
2&Attractive& stable\\
\hline
$\ge3$&Repellent& not stable\\
\hline\hline
\end{tabular}
\end{table}

To provide some intuition for our results, we illustrate numerically the non-stability of Kyle's equilibrium when $N=3$. We numerically  compute the limit of the policy iterations when the insider’s measurement of noise-trader variance slightly differ from the correct noise-trader variance. Our numerics are based on a 10th digit difference. Then, we numerically illustrate that the policy iterations converge, however, not to Kyle's equilibrium. This happens because there is an eigenvalue of the Jacobian of the policy iteration operator evaluated at Kyle's equilibrium strictly bigger than one (in absolute value), which implies non- stability. However, there are fixed points for which all corresponding eigenvalues are strictly less than one (in absolute value). Whenever the insider's variance estimate differs from the correct variance, the policy iterations converge numerically to such a fixed point.

%Third,  also when $N\ge3$, we show that the degree of non stability of Kyle's equilibrium is not severe enough to cause Kyle's equilibrium to being unstable. Mathematically speaking, for $N \in \{1,2\}$, we show that Kyle's equilibrium is an attracting fixed point for the policy iteration operator whereas, for $N\ge3$, we show that the fixed point is repelling. 

While the main part of the paper is about policy iterations in the insider's control (i.e., informed stock orders), we consider an alternative in Appendix \ref{sec_alternative}, where we iterate  the market makers' control (i.e., the pricing rule). This variation leads to the same stability conclusions in that Kyle's equilibrium is stable for $N\in \{1,2\}$ and not stable for $N\ge3$.

All proofs are in Appendix \ref{app_proofs}. Our proofs rely on a new characterization of Kyle's equilibrium in terms of a one-dimensional  fully  autonomous recursion, which is independent of all model inputs. 

Throughout the text, we use the symbol $'$ to transpose vectors. For example, $\vec x = (x_1,...,x_N)'$ denotes a column vector in $\R^N$. For numbers, we use ... to indicate that we have excluded remaining decimals. For example, we have $\pi = 3.14159...$.

\section{Kyle's discrete-time model} This section briefly recalls the discrete-time model in Kyle (1985) with $N\in\N$ trading times. The noise traders' orders $\Delta u_n$ at trading time $n\in \{1,...,N\}$ are Gaussian random variables with mean zero and variance $\sigma_u^2 \Delta $, where $\Delta >0$  is the time step. The stock's  liquidating value is denoted by $\tv$, which is assumed  Gaussian with  mean zero and variance $\Sigma_0:= \V[\tv]>0$. These exogenous random variables $(\tv,\Delta u_1,...,\Delta u_N)$ are assumed mutually independent. 

At time $n=1$, the insider submits orders $\Delta x_1$ to the market makers. The orders $\Delta x_1$ are required to be measurable with respect to $\sigma(\tv)$. At later times $n\in \{2,...,N\} $, the insider submits orders $\Delta x_n$, which are required to be measurable with respect to $\sigma(\tv,\Delta u_1,...,\Delta u_{n-1})$.  The aggregate orders are defined as
\begin{align}\label{Dyn}
\Delta y_n := \Delta u_n + \Delta x_n,\quad n\in\{1,...,N\}.
\end{align}
For a given pricing rule $p_n = p_n(\Delta y_1,..., \Delta y_n)$, the insider seeks orders $(\Delta x_1,...,\Delta x_n)$ that maximize her expected profit given by
\begin{align}\label{objective}
\sum_{n=1}^N \E\big[(\tv-p_n)\Delta x_n\big| \tv \big].
\end{align}

The market makers set prices $p_n$ in the following sense. At time $n\in\{1,...,N\}$, the market makers observe the aggregate orders $\Delta y_n$ from \eqref{Dyn} before setting the stock price as 
\begin{align}\label{eff_price}
p_n = \E[\tv | \Delta y_1,...,\Delta y_n ],\quad n\in\{1,...,N\}.
\end{align}

The next result (due to Kyle, 1985) gives existence of a linear Kyle equilibrium in the sense that items 2. and 3.  in Theorem \ref{thm_Kyle} hold.

\begin{theorem}[Kyle, 1985]\label{thm_Kyle} 
\begin{enumerate} 
\item For $\Delta>0, \sigma_u>0, \Sigma_0:=\V[\tv] >0$, and $\alpha_N:=0$, there exist unique solutions $(\hat \lambda_n,\hat \Sigma_n,\hat \alpha_n,\hat \beta_n)$, $n=1,...,N$, of 
\begin{align}\label{Kyleformulas}
\begin{split}
\lambda_n &= \frac{\beta_n\Sigma_{n-1}}{\beta_n^2\Sigma_{n-1}\Delta+\sigma_u^2},\quad \Sigma_n = \frac{\Sigma_{n-1}\sigma_u^2}{\beta_n^2\Sigma_{n-1}\Delta+\sigma_u^2},\\
\alpha_{n-1} &= \frac1{4\lambda_n(1-\alpha_n\lambda_n)},\quad \beta_n = \frac{1-2\alpha_n\lambda_n}{\Delta2\lambda_n(1-\alpha_n\lambda_n)},\quad n=1,...,N,
\end{split}
\end{align}
such that the second-order condition $\lambda_n\alpha_n < 1$ holds.

\item For the pricing rule
\begin{align}\label{Dpn}
\Delta p_n := \hat \lambda_n \big(\Delta u_n + \Delta x_n\big),\quad p_0 := 0,
\end{align}
the insider's optimal orders are
\begin{align}\label{Dxn}
\begin{cases}
\Delta \hat x_n = \hat \beta_n(\tv - \hat p_{n-1})\Delta,\\
\Delta \hat p_n = \hat \lambda_n \big(\Delta u_n + \Delta \hat x_n\big),\quad \hat p_0=0.
\end{cases}
\end{align}

\item For the orders \eqref{Dxn}, the stock price \eqref{Dpn} is efficient in the sense that \eqref{eff_price} holds.

\end{enumerate}
\end{theorem}
In what follows, we will refer to $\vec{\hat{\beta}}$ as the insider's equilibrium trading intensity as $\vec{\hat{\beta}}$ determines how aggressively the insider trades when the market price differs from her own valuation.

\section{Policy iterations and stability}
To iteratively create a sequence of insider orders, we start with some vector $\vec \beta^{(0)}=(\beta_1^{(0)},...,\beta_N^{(0)})'\in (0,\infty)^N$, which differs only marginally from Kyle's equilibrium $\vec{\hat \beta}$ from Theorem \ref{thm_Kyle}. Kyle's lambda $\vec \lambda^{(0)}=(\lambda_1^{(0)},...,\lambda_N^{(0)})'\in \R^N$ is defined similarly to \eqref{Kyleformulas}  by $\Sigma^{(0)}_0 :=\V[\tv]>0$ and
\begin{align}\label{Dxniterative1}
\lambda^{(0)}_n &:= \frac{\beta^{(0)}_n\Sigma^{(0)}_{n-1}}{\big(\beta_n^{(0)}\big)^2\Sigma^{(0)}_{n-1}\Delta+\sigma_u^2},\quad \Sigma^{(0)}_n := \frac{\Sigma^{(0)}_{n-1}\sigma_u^2}{\big(\beta^{(0)}_n\big)^2\Sigma^{(0)}_{n-1}\Delta+\sigma_u^2},\quad n=1,...,N.
\end{align}
The initial pricing rule is defined by
\begin{align}\label{initial_pricing_rule}
\Delta p^{(0)}_n := \lambda^{(0)}_n \big(\Delta u_n + \Delta x_n\big),\quad p^{(0)}_0 := 0,
\end{align}
where ($\Delta x_1,...,\Delta x_N)$ denote arbitrary insider orders. When she faces the pricing rule \eqref{initial_pricing_rule}, the insider's optimal orders that maximize \eqref{objective} are similar to \eqref{Dxn}, and given by
\begin{align}\label{Dxniterative}
\begin{cases}
\Delta x^{(1)}_n := \beta^{(1)}_n(\tv - \hat{p}^{(0)}_{n-1})\Delta,\\
\Delta \hat {p}^{(0)}_n := \lambda^{(0)}_n \big(\Delta u_n + \Delta x^{(1)}_n\big),\quad \hat p^{(0)}_0 := 0.
\end{cases}
\end{align}
In \eqref{Dxniterative}, the next policy iteration $\vec \beta^{(1)}:=(\beta_1^{(1)},...,\beta_N^{(1)})'$  is computed by 
 $\alpha^{(1)}_N :=0$ and
\begin{align}\label{beta_iteration}
\beta^{(1)}_n := \frac{1-2\alpha^{(1)}_n\lambda^{(0)}_n}{\Delta2\lambda^{(0)}_n(1-\alpha^{(1)}_n\lambda^{(0)}_n)},\quad \alpha^{(1)}_{n-1} &:= \frac1{4\lambda^{(0)}_n(1-\alpha^{(1)}_n\lambda^{(0)}_n)},\quad n=N,...,1.
\end{align}
Given the pricing rule \eqref{initial_pricing_rule}, the orders \eqref{Dxniterative} maximize \eqref{objective} provided that the second-order condition $\alpha^{(1)}_n\lambda^{(0)}_n < 1$ holds. However, because Kyle's equilibrium coefficients  from Theorem \ref{thm_Kyle} satisfy $\hat \alpha_n\hat \lambda_n < 1$, a continuity argument gives that $\alpha^{(1)}_n\lambda^{(0)}_n < 1$ provided that $|\vec \beta^{(0)}-\vec{\hat \beta}|<\epsilon$ for some $\epsilon >0$ sufficiently small. Here, $|\vec \beta|$ denotes the standard Euclidean norm of $\vec \beta\in \R^N$ given by $|\vec \beta|:= \sqrt{\sum_{n=1}^N \beta_n^2}$. 

We write the above policy iteration step compactly as 
\begin{align}\label{Kyle_FP}
\vec \beta^{(1)} = T(\vec \beta^{(0)}),
\end{align}
for a non-linear smooth function $T:\R^N\to\R^N$ with domain  dom$(T)\subset \R^N$. For $\vec \beta \notin \text{dom}(T)$, we set $T(\vec \beta) := (\infty,...,\infty)'$. Of course, Kyle's equilibrium coefficients $\vec{\hat \beta} = (\hat \beta_1,...,\hat \beta_N)'$  from Theorem \ref{thm_Kyle} satisfy the fixed-point property
\begin{align}\label{eq_cond1}
 \vec{\hat \beta}=T(\vec{\hat \beta}),
\end{align}
but there are several other solutions to \eqref{eq_cond1}. The $n$'th coordinate of the function $T$ in \eqref{Kyle_FP} is given as a ratio of polynomial functions (i.e., $T$ is a \emph{rational function}). Because the general expression for $T$ is long and not needed in our stability analysis for $N\ge3$, we only give  $T$ for $N\in \{1,2\}$.  

\begin{example}\label{ex_Nequls2} \begin{enumerate}
\item For $N=1$, the function $T$ in \eqref{Kyle_FP} is given by
\begin{align}\label{Twhenn=1}
T(\beta) := \frac{\beta^2\Sigma_0\Delta +\sigma_u^2}{2\Delta \beta\Sigma_0}.
\end{align}
The domain of $T$ is given by
$$
\text{dom}(T)= \{ \beta\in  \R: \beta\neq 0\}.
$$
\item For $N=2$, the function $T$ in \eqref{Kyle_FP} is given by
\begin{align}\label{T2}
T \begin{pmatrix}
\beta_1\\
\beta_2
\end{pmatrix}&=
\begin{pmatrix}
\frac{\left(\beta_1^2 \Delta  \Sigma_0+\sigma_u^2\right) \left(\beta_1 \Delta  \Sigma_0 (\beta_1-\beta_2)^2+\sigma_u^2 (\beta_1-2 \beta_2)\right)}{\beta_1 \Delta  \Sigma_0 \left(\beta_1 \Delta  \Sigma_0 \left(\beta_1^2-4 \beta_1 \beta_2+\beta_2^2\right)+\sigma_u^2 (\beta_1-4 \beta_2)\right)}\\
\frac{\Delta  \Sigma_0 \left(\beta_1^2+\beta_2^2\right)+\sigma_u^2}{2 \beta_2 \Delta  \Sigma_0}
\end{pmatrix}.
\end{align}
The domain of $T$ is given by
\begin{align*}
\text{dom}(T)= \{ (\beta_1&,\beta_2)'\in  \R^2 : \\
&\beta_1 \left(\beta_1 \Delta  \Sigma_0 \left(\beta_1^2-4 \beta_1 \beta_2+\beta_2^2\right)+\sigma_u^2 (\beta_1-4 \beta_2)\right)\neq 0, \beta_2\neq 0\}.
\end{align*}
\end{enumerate}
$\endproof$
\end{example}

Based on $\vec \beta^{(1)}$ from \eqref{Kyle_FP}, we use forward recursion to iteratively construct the sequence $\vec \beta^{(2)}, \vec \beta^{(3)}, ...$. More specifically, given the $m$'th policy iteration $\vec \beta^{(m)}\in \R^N$, the next policy iteration is defined as 
\begin{align}\label{Kyle_FPm}
\vec \beta^{(m+1)}:= T(\vec \beta^{(m)}),\quad m\in \{0,1,...\}.
\end{align}

We use the following definition of stability, which is based on Definition 1.3 in S\"uli and Mayers (2003).

\begin{definition}\label{Def_eq}
Kyle's equilibrium is  \emph{locally stable} with respect to policy iterations for the  insider if there exists $\epsilon >0$ such that all initial policies $\vec \beta^{(0)}\in\text{dom}(T)$ with $0<|\vec \beta^{(0)} -\vec{\hat \beta} |<\epsilon$ satisfy
\begin{align}\label{stab1}
\lim_{m\to \infty} \vec \beta^{(m)} = \vec{\hat \beta},
\end{align}
where the sequence $\vec \beta^{(m)}$ is defined recursively by \eqref{Kyle_FPm}.
$\endproof$
\end{definition}
\noindent In Definition \ref{Def_eq}, the term \emph{locally} refers to the smallness condition $|\vec \beta^{(0)} -\vec{\hat \beta} |<\epsilon$. % and the term \emph{linearly} refers to the starting policy $\Delta x^{(0)}_n$ having the particular form in \eqref{Dxniterative} for $\vec \beta^{(0)}\in \R^N$.  
Definition \ref{Def_eq2} below allows for more general policy iterations.

\begin{figure}[h]
	\includegraphics[scale=0.5]{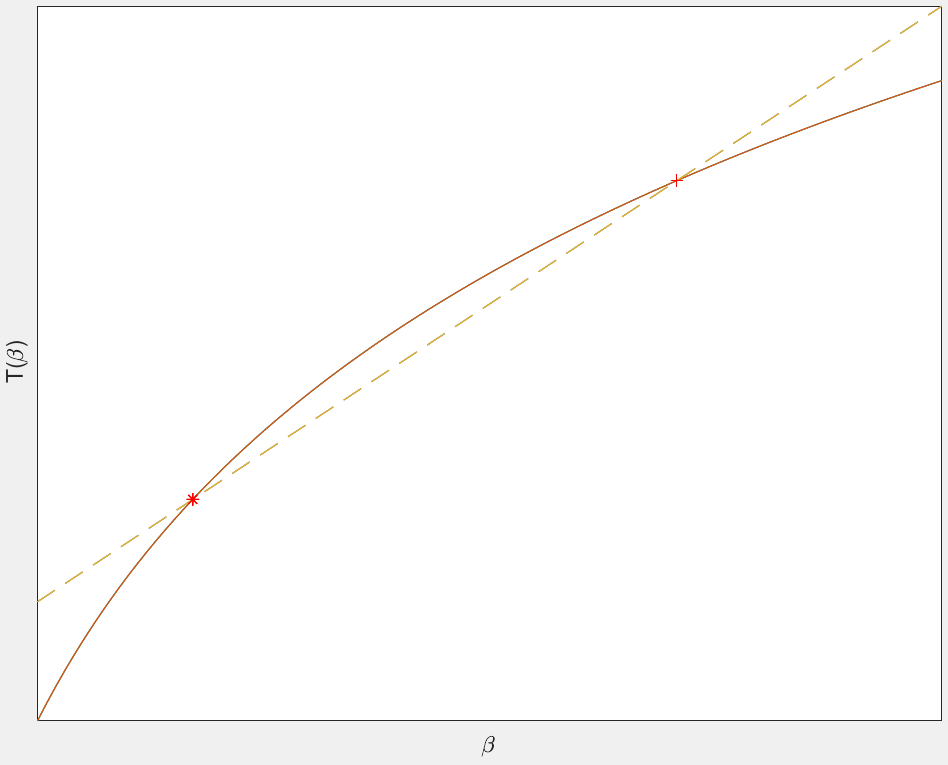}
		\centering
		\caption{Two fixed points of a fictitious operator $T:\R\to\R$. The left fixed point is not stable and the right fixed point is stable.}
	\label{f:stability}
\end{figure}

To gain some intuition of local stability,  Figure \ref{f:stability} depicts a fictitious operator $T:\R\to\R$ with two fixed points. The dashed line is the 45-degree line and the intersections of the two lines correspond to $T$'s two fixed points.  The fixed point to the left is not stable because policy iterations starting from a vicinity of this point moves away from this fixed point. This is called a \emph{repellent} fixed point. At the left fixed point, the graph of $T$ intersects the 45-degree line from below at this point indicating that $T$'s derivative is larger than $1$. On the other hand, the fixed point to the right is stable and $T$'s derivative at this point is smaller than $1$. This is called an \emph{attractive} fixed point. In the subsequent sections, we prove that the derivative of the policy iteration operator $T$ in \eqref{Kyle_FP}  evaluated at $\vec{\hat \beta}$ has a norm less than one when $N\in\{1,2\}$, and a norm larger than $1$ when $N\ge3$. In view of the above discussion, these norms indicate that Kyle's equilibrium is stable if and only if $N\in\{1,2\}$. 

\section{One or two trading times }
Let $T$ be defined in \eqref{Twhenn=1} for $N=1$ or \eqref{T2} for $N=2$. In the next result, $\nabla T(\vec \beta)\in \R^{N\times N}$ denotes the Jacobian matrix of $T$'s derivatives evaluated at $\vec \beta\in \text{dom}(T)$.
\begin{theorem}\label{lemN2} For $\Delta >0,\,\sigma_u>0$, and $\Sigma_0>0$, we have:
\begin{enumerate}
\item For $N:=2$, the Jacobian matrix of $T$'s derivative satisfies
\begin{align}\label{Jacob1}
\nabla T (\vec{
\hat \beta})&=
\begin{pmatrix}
-0.981214... \quad & 0\\
0.554958... \quad  &0
\end{pmatrix}.
\end{align}
\item  For $N\in \{1,2\}$, Kyle's equilibrium is locally stable with respect to policy iterations for the insider in the sense of Definition \ref{Def_eq}.
\end{enumerate}
\end{theorem}
\noindent Our proof of Theorem \ref{lemN2}  shows that $T$ defined in \eqref{Twhenn=1} or \eqref{T2} is a local contraction for $\vec \beta$ near Kyle's equilibrium $\vec{\hat \beta}$ when $\R^N$ is equipped with the norm $|\vec\beta|_\infty:= \max\{|\beta_1|,...,|\beta_N|\}$ and when $N\in \{1,2\}$. For $N=1$, the proof gives the stronger property $T'(\hat \beta)=0$, which implies that Kyle's equilibrium is a super-attractive fixed point. Hubbard and Papadopol (1994) has details about super-attractive fixed-points. A similar observation is in  Boulatov and Bernhardt (2015).

\section{Three or more trading times}
 Before  rigorously proving that the policy iterations are not stable, we consider a numerical  illustration. 
\subsection{Numerics}\label{sec_num}
The recursive formulas in \eqref{Kyleformulas} in Theorem \ref{thm_Kyle} show that Kyle's equilibrium $\vec{\hat \beta}$ depends continuously on the input parameters  $\Delta>0$, $\sigma_u>0,$ and $\Sigma_0:=\V[\tv] >0$. This implies that for any $\epsilon>0$, we can choose $\delta>0$ such that the equilibrium $\vec \beta^{(0)}$ corresponding to a marginal perturbation of the noise-trader variance $\sigma^2_u \pm \delta$ satisfies $|\vec{\hat \beta} - \vec \beta^{(0)}|<\epsilon$, where  $\vec{\hat \beta}$ corresponds to the the equilibrium trading intensity of the insider when the noise-trader variance equals $\sigma^2_u$. To illustrate the non-stability of Kyle's equilibrium when $N=3$ and $\sigma_u := \Delta := \V[\tv]:=1$, we consider a 10th digit perturbation $\delta := \frac{1}{10^{10}}$ of the noise-trader variance. Kyle's equilibrium corresponding to the noise-trader variance $\sigma^2_u+ \delta = 1.0000000001$ is given by
\begin{align}\label{beta0}
	\vec {\beta}^{(0)}
	:=
	\begin{pmatrix}
		0.5381695932490208...\\ 
		0.7575868210661554... \\
		1.3651242810275332...
	\end{pmatrix}.
\end{align}
Kyle's equilibrium $\vec{\hat\beta}$ corresponding to the noise-trader variance $\sigma_u^2 = 1$ differs slightly from $\vec {\beta}^{(0)}$ and is given by
\begin{align}
	\vec {\hat \beta}
	:=
	\begin{pmatrix}
		0.5381695932221123...,\\
		0.7575868210282761...,\\
		1.3651242809592772...
	\end{pmatrix}.
\end{align}

To illustrate the non-stability, we start the policy iterations from \eqref{beta0} in a model with $\sigma_u:=\Delta := \V[\tv]:=1$. Numerically, the non-linear policy iterating scheme \eqref{Kyle_FPm} starting from  $\vec {\beta}^{(0)}$ converges to
\begin{align*}
	\vec {\beta}^{(\infty)}
	:=
	\begin{pmatrix}
		&1.2582536009629393...\\
		&-2.157491457005712...\\
		&2.6903478420808034...
	\end{pmatrix}\neq \vec { \hat {\beta}}.
\end{align*}

To gain some intuition for why $\vec \beta^{(m+1)} = T(\vec \beta^{(m)})=T\big(\cdots T(\vec \beta^{(0)})\big)$ converges to $\vec {\beta}^{(\infty)}$ instead of converging to Kyle's equilibrium $\vec { \hat {\beta}}$, Table \ref{tab_eigenvalues} reports two sets of eigenvalues.
\begin{table}[ht]
	\centering \caption{Eigenvalues for $\sigma_u = \V[\tv] =\Delta =1$ and $N=3$ }
	\label{tab1}
	\vspace{.2cm}
	\begin{tabular}{c|c}
		\hline
		$\nabla T(\vec  {\beta}^{(\infty)})$&\{0.413853..., 0.193926..., 0\}\\
		$\nabla T(\vec { \hat {\beta}})$&\{-2.16095..., -0.896373..., 0\}\\
		\hline
	\end{tabular}
	\label{tab_eigenvalues}
\end{table} \ \\
\noindent For an  induced matrix norm $||\cdot ||$, the value $||\nabla T(\vec { \hat {\beta}}) ||$ dominates (in absolute value) all of $\nabla T(\vec { \hat {\beta}})$'s eigenvalues. Therefore, the second row in Table \ref{tab_eigenvalues} shows that $||\nabla T(\vec { \hat {\beta}}) ||>1$, which indicates that Kyle's equilibrium is not stable. On the other hand, the eigenvalues for $\nabla T(\vec \beta^{(\infty)})$ in the first row in Table \ref{tab_eigenvalues} are smaller than one and therefore do not contradict local stability.  To build an analogy with Figure \ref{f:stability}, $\vec { \hat {\beta}}$ (resp. $\vec \beta^{(\infty)}$) can be associated with the non-stable left fixed point (resp. the stable right fixed point) at which the operator has a derivative bigger (resp. less)  than $1$.  Mathematically speaking, for $N =3$, Kyle's equilibrium  $\vec{\hat \beta}$ is a repelling fixed point for the policy iteration operator $T$ whereas the fixed point $\vec{\beta}^{(\infty)}$ is an attractive fixed point for $T$.

\subsection{Theory}
%The numerics in Section \ref{sec_num} below illustrate that when $N\ge3$, the gradient $\nabla T$ evaluated at Kyle's equilibrium $\vec{\hat{\beta}}$ has eigenvalues with absolute value strictly bigger than one.  This gives us the intuition that local stability fails and that  $\hat \beta_1$ is the most unstable variable, then comes $\hat \beta_2$, and so on whereas $\hat \beta_{N-1}$ and $\hat \beta_N$ are stable. This is the intuitive reason we focus on $\hat \beta_{N-2}$ in this section.

To rigorously disprove local stability when $N\ge3$, we iterate only in the third-to-last variable $\beta_{N-2}$ whereas all other coefficients are set equal to Kyle's equilibrium values. Then, we show that the resulting policy iterations diverge, and, consequently, there is no $\epsilon>0$ such that \eqref{stab1} holds. To this end, we let $T$ be from \eqref{Kyle_FP} and define the scalar function $\tilde T :\R \to\R$ as $T$'s $(N-2)$'th coordinate
\begin{align}\label{tildeT}
\tilde T(\beta^{(m)}_{N-2} ) :=  T\begin{pmatrix}
\hat \beta_1 \\
\hat \beta_2 \\
\vdots\\
\hat \beta_{N-3} \\
\beta^{(m)}_{N-2} \\
\hat \beta_{N-1} \\
\hat \beta_N 
\end{pmatrix}_{N-2},
\end{align}
where $(\hat \beta_1,...,\hat\beta_N)$ are Kyle's equilibrium coefficients from Theorem \ref{thm_Kyle} and $\beta^{(m)}_{N-2}\in \R$ is the variable we iterate in. When $\beta^{(m)}_{N-2}\in \R$ is such that $(\hat \beta_1,...,\hat \beta_{N-3},\beta^{(m)}_{N-2},\hat \beta_{N-1},\hat \beta_N)' \notin $ dom$(T)$, we set $\tilde T(\beta^{(m)}_{N-2} ):=\infty$.

\begin{theorem}\label{lemma_Main} Let $N\ge 3$. For $\Delta >0,\,\sigma_u>0$, and $\Sigma_0>0$, we have:
\begin{enumerate}
\item For any starting value $ \beta^{(0)}_{N-2} \in\R$ with $ \beta^{(0)}_{N-2} \neq \hat \beta_{N-2}$, the recursively defined sequence 
\begin{align}\label{lineariterationplus}
\beta^{(m+1)}_{N-2}  := \hat \beta_{N-2}+\tilde T'( \hat \beta_{N-2})\big(\beta^{(m)}_{N-2} - \hat \beta_{N-2}\big),\quad m=0,1,...,
\end{align}
diverges in the sense $\lim_{m\to\infty} |\beta^{(m)}_{N-2}| =\infty$.

\item  Kyle's equilibrium is not locally stable  with respect to policy iterations for the insider  in the sense of Definition \ref{Def_eq}.
\end{enumerate}
\end{theorem}

Theorem \ref{lemma_Main} implies that policy iterations based on more general starting policies are also not locally stable in the following sense. We say stock holdings $\vec x = (x_0,x_1,....,x_N)'\in \sL^2$ if the random variables $\Delta x_n :=x_n-x_{n-1}$ satisfy $||\vec x|| := \sqrt{\sum_{n=1}^N \E\big[(\Delta x_n)^2\big]}<\infty$.  An \emph{extension} $T^\circ$ of $T$ has  dom$(T)\subseteq$ dom$(T^\circ)\subseteq  \sL^2$ and is said to be \emph{consistent} with $T$ if $T^\circ(\vec x)=T(\vec x)$ for all $\vec x\in$ dom$(T)$.

\begin{definition}\label{Def_eq2}
Let $T^\circ$ be a consistent extension of $T$.  Kyle's equilibrium is  \emph{locally stable} with respect  to generalized policy iterations for the insider  if there exists $\epsilon >0$ such that all starting policies $\vec x^{(0)}\in \text{dom}(T^\circ)$ with $0< ||  \vec x^{(0)} -  \vec {\hat x}|| <\epsilon$ satisfy
\begin{align}\label{stab1a}
\lim_{m\to \infty} \vec x^{(m)} = \vec{\hat x},\quad  \vec x^{(m+1)} := T^\circ(\vec x^{(m)}),
\end{align}
where $T^\circ(\vec x):= (\infty,...,\infty)'$ whenever $x\notin \text{dom}(T^\circ)$.
$\endproof$
\end{definition}

The following by-product is an immediate consequence of Theorem \ref{lemma_Main}.

\begin{corollary} Let $N\ge 3$ and let $T^\circ$ be a consistent extension of $T$. Then, for $\Delta >0,\,\sigma_u>0$, and $\Sigma_0>0$, Kyle's equilibrium is not locally stable with respect  to generalized policy iterations for the insider  in the sense of Definition \ref{Def_eq2}.

\end{corollary}

We conclude this section by considering the related definition of an \emph{unstable} equilibrium. This definition can be found in, e.g.,  Definition 1.3 in S\"uli and Mayers (2003) and differs from non locally stable equilibria (i.e., unstable and non-stable are different mathematical concepts).

\begin{definition}\label{Def_eq3} If $\vec\beta^{(0)}= \vec{\hat \beta}$ is the only starting policy for which \eqref{stab1} holds, we say that Kyle's equilibrium is \emph{unstable} with respect to policy iterations for the  insider.

$\endproof$
\end{definition}
\noindent By comparing Definitions \ref{Def_eq} and \ref{Def_eq3}, we see that an unstable fixed point is always also not stable. However, Exercise 1.2 in S\"uli and Mayers (2003) shows that a non-stable fixed point can fail to be unstable.  For $N\ge 3$, our next and last theoretical result shows that while Kyle's equilibrium is not stable, it is also  not unstable. 

\begin{theorem}\label{thm3} Let $N\ge 3$. For $\Delta >0,\,\sigma_u>0$, and $\Sigma_0>0$, Kyle's equilibrium is not unstable with respect  to policy iterations for the insider  in the sense of Definition \ref{Def_eq3}.
\end{theorem}
\noindent In our proof of Theorem \ref{thm3}, we marginally  perturb the last equilibrium coordinate by setting $\beta^{(0)}_N:=\hat \beta_N+\delta$ for a small $\delta>0$ and $\beta^{(0)}_n :=\hat \beta_n$ for $n\in\{1,...,N-1\}$. Then, we show the corresponding policy iterations converge to $\vec{\hat \beta}$. Alternatively, the iterations also converge  to $\vec{\hat \beta}$ when we  set $\beta^{(0)}_{N-1}:=\hat \beta_{N-1}+\delta$  and $\beta^{(0)}_n :=\hat \beta_n$ for $n\in\{1,...,N-2,N\}$. However, no matter how small a perturbation, as soon as we perturb one of the first $N-2$ coordinates of  $\vec{\hat \beta}$, the policy iterations do not converge to $\vec{\hat \beta}$.

\section{Conclusion}
Based on a standard notion of stability used widely in both numerical analysis and  financial economics, we proved that the dynamic equilibrium model of informed trading in Kyle (1985) is stable when $N\in \{1,2\}$ and not stable when $N\ge3$. To investigate further the severity of non-stability, we proved that Kyle's equilibrium is not unstable when $N\ge3$. We numerically illustrated that policy iterations can converge to fixed points, which are not equilibria.

\appendix

\section{Policy iterations for the market markers}\label{sec_alternative}

We outline how to perform policy iterations in the market makers' control (i.e., how to iterate the pricing rule). Because this alternative policy iteration scheme produces the same conclusions as iterating the insider's orders, we keep the presentation brief.  

Given $\vec \lambda^{(m)}=(\lambda_1^{(m)},...,\lambda_N^{(m)})'\in \R^N$, we define $\vec \lambda^{(m+1)}=(\lambda_1^{(m+1)},...,\lambda_N^{(m+1)})'\in\R^N$ recursively as follows. We define the insider's response by starting from $n=N$ and going to $n=1$ by setting $\alpha^{(m)}_N :=0$ and
\begin{align*}
\beta^{(m)}_n = \frac{1-2\alpha^{(m)}_n\lambda^{(m)}_n}{\Delta2\lambda^{(m)}_n(1-\alpha^{(m)}_n\lambda^{(m)}_n)},\quad \alpha^{(m)}_{n-1} &= \frac1{4\lambda^{(m)}_n(1-\alpha^{(m)}_n\lambda^{(m)}_n)},\quad n=N,...,1.
\end{align*}

To update the markets makers' response, we start with $n=1$ and go to $n=N$ by defining $\Sigma_0^{(m+1)}:=\V[\tv]>0$ and
\begin{align*}
\lambda^{(m+1)}_n &:= \frac{\beta^{(m)}_n\Sigma^{(m+1)}_{n-1}}{\big(\beta_n^{(m)}\big)^2\Sigma^{(m+1)}_{n-1}\Delta+\sigma_u^2},\quad \Sigma^{(m+1)}_n := \frac{\Sigma^{(m+1)}_{n-1}\sigma_u^2}{\big(\beta^{(m)}_n\big)^2\Sigma^{(m+1)}_{n-1}\Delta+\sigma_u^2},\quad n=1,...,N.
\end{align*}
Similar to \eqref{Kyle_FPm}, we write this iteration compactly as
\begin{align}\label{Kyle_FPS}
\vec \lambda^{(m+1)} = S(\vec \lambda^{(m)}),
\end{align}
for a non-linear but smooth function $S:\R^N\to \R^N$. 

By replacing $T$ and $T(\vec\beta^{(m)})$ with $S$ and $S(\vec\lambda^{(m)})$ in Definitions \ref{Def_eq}, \ref{Def_eq2}, and \ref{Def_eq3}, we obtain the precise mathematical meaning of stability and instability with respect to policy iterations for the market makers. 
\begin{theorem}\label{thm:appendix}For $\Delta >0,\,\sigma_u>0$, and $\Sigma_0>0$, we have:
\begin{enumerate}
\item For $N\in \{1,2\}$, Kyle's equilibrium is locally stable with respect to policy iterations for the market makers. 
\item For $N\ge 3$, Kyle's equilibrium is not locally stable with respect to policy iterations for the market makers.
\item For $N\ge 3$, Kyle's equilibrium is not unstable with respect to policy iterations for the market makers.
\end{enumerate}
\end{theorem}
\noindent We omit the proofs because they are similar to the proofs of Theorems  \ref{lemN2}, \ref{lemma_Main}, and \ref{thm3}. %Alternatively, we can argue by contradiction. For example, if we for item 3. in Theorem \ref{thm:appendix}  assume there exists  $\vec \lambda^{(0)}\in \R^N$ such that $\vec \lambda^{(0)}\neq \vec{\hat{\lambda}}$ and $\vec \lambda^{(m+1)} = S\big(\cdots S(\vec \lambda^{(0)})\big)$ converges to $ \vec{\hat{\lambda}}$, then $\vec\beta^{(0)} := $

\section{Proofs} \label{app_proofs}

\subsection{Autonomous recursion}

We start by providing an autonomous recursion, which will be used in our proofs. A related recursion appears in Proposition 2 in Holden and Subrahmanyan (1992). Instead of Kyle's equilibrium coefficients $(\hat \beta_1,...,\hat\beta_N)$, we will use the coefficients $(\hat b_1,..,\hat b_N)$ in the next lemma. We prefer  $(\hat b_1,..,\hat b_N)$ because they are independent of the model input parameters $(\Delta, \sigma_u, \V[\tv])$.

\begin{lemma} \label{lem_auto}
\begin{enumerate}
\item There exist unique coefficients $(\hat b_n)_{n=1}^{N-1}\subset (0,1)$ given by the backward recursion
\begin{align}\label{b_recursion}
 \hat b_N=1,\quad \hat b_{n}^2 = \frac{\hat b_{n-1}^2}{(1-\hat b_{n-1}^2)^2(1+\hat b_{n-1}^2)},\quad n=N,N-1,...,1.
\end{align}
\item Kyle's equilibrium coefficients \eqref{Kyleformulas} can be expressed as
\begin{align}\label{betainsteadofb}
\hat \beta_n=\frac{\hat b_n\sigma_u}{\sqrt{\hat\Sigma_{n-1}\Delta}},\quad \hat\Sigma_n 
= \frac{\hat\Sigma_{n-1}}{1+\hat b_n^2},\quad \hat\Sigma_0 :=\V[\tv]>0,\quad n=1,2,...,N.
\end{align}
\end{enumerate}

\end{lemma}
\proof 1. We proceed by backward induction and assume $\hat b_n\in (0,1]$. The transformation $a := \hat b_{n-1}^2$ produces the third-degree polynomial
\begin{align}\label{b_recursion2}
\hat b_{n}^2 = \frac{a}{(1-a)^2(1+a)}.
\end{align}
For $\hat b_n\in (0,1]$, the polynomial \eqref{b_recursion2} has exactly one root $a\in (0,1)$. \ \\

\noindent 2. Inserting $\hat \beta_n=\frac{\hat b_n\sigma_u}{\sqrt{\hat\Sigma_{n-1}\Delta}}$ into the formula for $\Sigma_n$ in \eqref{Kyleformulas} gives $\hat\Sigma_n 
= \frac{\hat\Sigma_{n-1}}{1+\hat b_n^2}$. Solving for $\alpha_n$ in the formula for $\beta_n$ in \eqref{Kyleformulas} produces
\begin{align}\label{newalpha}
\begin{split}
\hat \alpha_n = \frac{2\hat \beta_n\Delta\hat\lambda_n-1}{2\hat\lambda_n(\hat\beta_n\Delta \hat\lambda_n-1)}=\frac{\sqrt{(\hat b_n^2+1) \Delta  \hat\Sigma_n}-2 \hat b_n \Delta  \hat\lambda_n\sigma_u}{2 \hat \lambda_n \sqrt{(\hat b_n^2+1) \Delta  \hat\Sigma_n}-2 \hat b_n \Delta  \hat\lambda_n^2\sigma_u}.
\end{split}
\end{align}
From Eq. (3.18) in Kyle (1985), we have $\hat\lambda_n=\frac{\hat\beta_n \hat\Sigma_n}{\sigma_u^2}$. Then, the recursion for $\alpha_n$ in \eqref{Kyleformulas} becomes the recursion in \eqref{b_recursion}.

Finally, the terminal condition $\hat b_N=1$ in \eqref{b_recursion} comes from $\hat\alpha_N=0$. This is because \eqref{newalpha} for $n=N$ gives the requirement
$$
1= 2\hat \beta_N\Delta\hat\lambda_N = \frac{2\hat b_N^2}{1+\hat b_N^2},
$$
which has  $\hat b_N=1$ as its only positive solution.
$\endproof$

\subsection{Mathematical proofs}

For a matrix $A\in \R^{N\times N}$, we recall the matrix norm $||A||_\infty := \max_{i\in\{1,...,N\}} \sum_{j=1}^N |A_{ij}|$. The matrix norm $||\cdot ||_\infty$ is  induced by the vector norm $|\vec\beta|_\infty := \max\{|\beta_1|,...,|\beta_N|\}$ and therefore the inequality $|A\vec \beta|_\infty \le ||A||_\infty |\vec\beta|_\infty$ holds. The second part of the following proof is standard and can be found in, e.g., 
the proof of Theorem 4.2 in S\"uli and Mayers (2003).

\proof[Proof of Theorem \ref{lemN2}] 1. We consider the parametrization
$$
\beta_1 := \frac{b_1\sigma_u}{\sqrt{\Sigma_0 \Delta}},\quad \beta_2:= \frac{b_2\sigma_u}{\sqrt{\Sigma_1 \Delta}},\quad b_1,b_2\in\R.
$$
Lemma \ref{lem_auto} ensures $(b_1,b_2):= (\hat b_1,\hat b_2)$ produces Kyle's equilibrium. Furthermore, we can use \eqref{T2} to calculate $\nabla T (\vec\beta)$ for $\vec \beta = (\beta_1,\beta_2)'$. 
Inserting Kyle's equilibrium $\vec {\hat \beta}$ into $\nabla T$ gives \eqref{Jacob1}.

2.  First, we consider $N=1$. The policy iteration function $T$ in \eqref{Kyle_FP} is given in  \eqref{Twhenn=1}. Our argument is based on the second-order Taylor expansion around  Kyle's equilibrium $\hat \beta:=\frac{\sigma_u}{\sqrt{\Delta \Sigma_0}}$ given by
\begin{align}\label{TaylorN1}
\begin{split}
T(\beta) &= T(\hat \beta)+ T'(\hat \beta)\big(\beta -\hat \beta\big)+ \frac12T''(\gamma )\big(\beta -\hat \beta\big)^2\\
&= \hat \beta+ \frac12\frac{\sigma_u^2}{\gamma^3\Delta \Sigma_0}\big(\beta -\hat \beta\big)^2,
\end{split}
\end{align}
where $\gamma$ is a point between $\beta$ and $\hat\beta$. The second equality in \eqref{TaylorN1} follows from $T'(\hat \beta)= 0$.\footnote{For $N=1$, the property $T'(\hat \beta)= 0$ implies that $\hat \beta$ is a super-attracting fixed point and implies that the policy iterations converge faster than linearly.}

Let $\epsilon \in (0,1)$ be such that (i) $\epsilon <\hat \beta$ and (ii) $\frac{\epsilon\sigma_u^2}{(\hat\beta -\epsilon)^3\Delta \Sigma_0}<2$. Let $\beta^{(0)}>0$ with $0<|\beta^{(0)}-\hat \beta|<\epsilon$ be arbitrary. By combining the expressions in \eqref{Kyle_FPm} and \eqref{TaylorN1}, we get
\begin{align*}
\big|\beta^{(1)}-\hat\beta\big|&\le \frac12\frac{\sigma_u^2}{(\hat\beta -\epsilon)^3\Delta \Sigma_0}\big(\beta^{(0)} -\hat \beta\big)^2\\
&\le \frac12\frac{\epsilon\sigma_u^2}{(\hat\beta -\epsilon)^3\Delta \Sigma_0}\big|\beta^{(0)} -\hat \beta\big|.
\end{align*}
Iterating this inequality forward gives
\begin{align*}
\big|\beta^{(m+1)}-\hat\beta\big|&\le  \bigg(\frac12\frac{\epsilon\sigma_u^2}{(\hat\beta -\epsilon)^3\Delta \Sigma_0}\bigg)^{m+1}\big|\beta^{(0)} -\hat \beta\big|,\quad m\in \N,
\end{align*}
which converges to zero as $m\to \infty$.

Second, we consider $N=2$.  Based on  \eqref{Jacob1}, we have $||\nabla T(\vec{\hat \beta})||_\infty=0.981214...$. Because the function $\text{dom}(T)\ni\vec\beta \to \nabla T(\vec \beta)_{ij}$ is continuous for each matrix entry $i,j \in\{1,...,N\}$,  we have continuity of $\text{dom}(T)\ni\vec\beta \to ||\nabla T(\vec \beta)||_\infty$. Therefore, there exists an $\epsilon >0$ such that 
$$
|| \nabla T(\vec \beta)||_\infty<0.99,\quad \text{ whenever }\quad |\vec\beta -\vec{\hat \beta}|_\infty<\epsilon.
$$
For $|\vec\beta^{(m)} -\vec{\hat \beta}|_\infty<\epsilon$ and $t\in[0,1]$, we have 
$$
\big|\big(\vec \beta^{(m)} +t(\vec{\hat \beta}-\vec\beta^{(m)})\big)-\vec{\hat \beta}\big|_\infty = (1-t)|\vec \beta^{(m)}-\vec{\hat \beta}|_\infty<\epsilon,
$$
and therefore $ || \nabla T\big(\vec \beta^{(m)} +t(\vec{\hat \beta}-\vec\beta^{(m)})\big)||_\infty<0.99$. 
The fundamental theorem for line integrals gives the representation
\begin{align*}
\vec{\hat \beta} -\vec\beta^{(m+1)}  &= T(\vec{\hat \beta}) -T(\vec\beta^{(m)})  \\
&= \int_0^1  \nabla T\big(\vec \beta^{(m)} +t(\vec{\hat \beta}-\vec\beta^{(m)})\big)(\vec{\hat \beta}-\vec\beta^{(m)})dt.
\end{align*}
Applying $|\cdot |_\infty$ produces the inequality
\begin{align*}
|\vec{\hat \beta} -\vec\beta^{(m+1)}|_\infty &\le \int_0^1 || \nabla T\big(\vec \beta^{(m)} +t(\vec{\hat \beta}-\vec\beta^{(m)})\big)||_\infty|\vec{\hat \beta}-\vec\beta^{(m)}|_\infty dt\\
&\le0.99|\vec{\hat \beta}-\vec\beta^{(m)}|_\infty.
\end{align*}
By iterating this inequality forward,  we see that 
$$
|\vec{\hat \beta} -\vec\beta^{(m+1)}|_\infty \le 0.99^{m+1}|\vec{\hat \beta}-\vec\beta^{(0)}|_\infty,\quad m\in \N,
$$ 
which converges to zero as $m\to \infty$. 

$\endproof$

\proof[Proof of Theorem \ref{lemma_Main}] 

1. {\bf Step 1/4:}  We start by rewriting \eqref{tildeT} as
\begin{align}\label{foverg}
\tilde T(\beta_{N-2}) = \frac{f(\vec{ B})}{g(\vec B)},\quad \vec B := (\hat \beta_1,...,\hat \beta_{N-3},\beta_{N-2},\hat \beta_{N-1}, \hat \beta_N )' ,
\end{align}
for $\beta_{N-2}\in \R$ such that $\vec B \in \text{dom}(T) = \{ \vec\beta\in \R^N: g(\vec \beta )\neq 0\}$. In \eqref{foverg},  the polynomials $f,g:\R^N \to \R$ are defined as
\begin{align}\label{fandg}
\begin{split}
f(\vec\beta) & := \big(\Delta  \Sigma_0(\beta_1^2+...+\beta_{N-2}^2)+\sigma ^2\big)\times\\ &\Big(\beta_{N-1}^2 \big(\Delta  \Sigma_0(\beta_1^2+...+\beta_{N-2}^2)+\sigma ^2\big) \big(\Delta  \Sigma_0\left(\beta_1^2+...+\beta_{N-2}^2+4\beta_{N-2}\beta_N+\beta_{N}^2\right)+\sigma ^2\big)\\
&+\beta_{N-1}^4 \Delta  \Sigma_0\big(\Delta  \Sigma_0(\beta_1^2+...+\beta_{N-2}^2+2\beta_{N-2}\beta_N)+\sigma ^2\big)\\
&-4\beta_{N-1}^3\beta_N \Delta  \Sigma_0\big(\Delta  \Sigma_0(\beta_1^2+...+\beta_{N-2}^2)+\sigma ^2\big)-4\beta_{N-1}\beta_N \big(\Delta  \Sigma_0(\beta_1^2+...+\beta_{N-2}^2)+\sigma ^2\big)^2\\
&+2\beta_{N-2}\beta_N \big(\Delta  \Sigma_0(\beta_1^2+...+\beta_{N-2}^2)+\sigma ^2\big)^2\Big),\\
g(\vec\beta) & := 2 \beta_{N-2} \Delta  \Sigma_0 \bigg(-4 \beta_{N-1}^3 \beta_{N} \Delta  \Sigma_0 \big(\Delta  \Sigma_0(\beta_1^2+...+\beta_{N-2}^2)+\sigma ^2\big)\\
&+\beta_{N-1}^2 \big(\Delta  \Sigma_0 (\beta_1^2+...+\beta_{N-2}^2)+\sigma ^2\big) \Big(\Delta  \Sigma_0 \big(\beta_1^2+...+\beta_{N-3}^2+(\beta_{N-2}+\beta_{N})^2\big)+\sigma ^2\Big)\\
&-4 \beta_{N-1} \beta_{N} \big(\Delta  \Sigma_0 (\beta_1^2+...+\beta_{N-2}^2)+\sigma ^2\big)^2+\beta_{N-2} \beta_{N} \big(\Delta  \Sigma_0 (\beta_1^2+...+\beta_{N-2}^2)+\sigma ^2\big)^2\\
&+\beta_{N-1}^4 \Delta  \Sigma_0 \Big(\Delta  \Sigma_0 \big(\beta_1^2+...+\beta_{N-3}^2+\beta_{N-2} (\beta_{N-2}+\beta_{N})\big)+\sigma ^2\Big)\bigg),
\end{split}
\end{align}
for $\vec\beta = (\beta_1,...,\beta_N)'\in \R^N$. As functions of $\beta_{N-2}\in \R$ alone,  $f$ is a polynomial of degree 7 and $g$ is a polynomial of degree 6. \ \\

\noindent{\bf Step 2/4:} By substituting $\hat\beta_n$ from \eqref{betainsteadofb} for $n\neq N-2$, and $\beta_{N-2}= \frac{z \sigma_u}{\sqrt{\Delta \Sigma_0}}$ for $z>0$, we can find polynomials $\tilde{f},\tilde g:\R\to \R$ that are independent of $(\Delta, \sigma_u, \Sigma_0)$ such that \eqref{foverg} becomes
$$
\tilde T(\beta_{N-2}) = \tilde T(\frac{z \sigma_u}{\sqrt{\Delta \Sigma_0}}) = \frac{\frac{\sigma^8}{\Delta \Sigma_0} \tilde f(z)}{\frac{\sigma^7}{\sqrt{\Delta \Sigma_0}}\tilde g(z)}.
$$
The chain rule gives the derivative
$$
\tilde T'(\beta_{N-2}) = \frac{\sqrt{\Delta \Sigma_0}}{\sigma_u}\frac{\partial}{\partial z}\tilde T( \frac{z \sigma_u}{\sqrt{\Delta \Sigma_0}})= \frac{\partial}{\partial z}\frac{ \tilde f(z)}{\tilde g(z)}.
$$

\noindent{\bf Step 3/4:} Based on the previous step, we can assume $\Delta= \sigma_u= \Sigma_0=1$ without loss of generality. In this step, we eliminate the dependence on $N$ by reversing indices in \eqref{b_recursion}. That is, we let $\tilde b_n \in (0,1)$ be uniquely given by the forward recursion
\begin{align}\label{b_recursiontake2}
 \tilde b_1=1,\quad \tilde b_{n-1}^2 = \frac{\tilde b_{n}^2}{(1-\tilde b_{n}^2)^2(1+\tilde b_{n}^2)},\quad n=2,...
\end{align}
Unlike $(\hat b_1,\hat b_2,...)$ from \eqref{b_recursion}, the values $(\tilde b_1,\tilde b_2,...)$ in \eqref{b_recursiontake2} do not depend on $N\ge3$.

We augment $\vec {\hat \beta}$ and write $\vec {\hat \beta}_N=\big(\hat \beta_{N,1},...,\hat \beta_{N,N}\big)'\in \R^N$ to highlight its dependence on $N$. Similarly, we augment notation and write $(\tilde T_N, f_N, g_N)$ instead of $(\tilde T,f,g)$. We use \eqref{betainsteadofb} to write $\vec {\hat \beta}_N$ as
\begin{align*}
\hat \beta_{N,1} &=  \tilde{b}_N,\quad \hat \beta_{N,2} =  \tilde{b}_{N-1}\sqrt{1+ \tilde{b}_N^2},\quad \hat \beta_{N,3} =  \tilde{b}_{N-2}\sqrt{1+ \tilde{b}_N^2}\sqrt{1+ \tilde{b}_{N-1}^2},\quad ...,\\
\hat \beta_{N,N} &=  \tilde{b}_{1}\sqrt{1+ \tilde{b}_N^2}...\sqrt{1+ \tilde{b}_2^2}.
\end{align*}
We insert these expressions for $\hat \beta_{N,1},...,\hat \beta_{N,N}$ into \eqref{fandg} and use
\begin{align*}
\tilde T_N'(\hat \beta_{N,N-2}) &= \frac{f_N'(\hat \beta_{N,N-2})g_N(\hat \beta_{N,N-2})-f_N(\hat \beta_{N,N-2})g'_N(\hat \beta_{N,N-2})}{g_N(\hat \beta_{N,N-2})^2},\\
\tilde T_{N-1}'(\hat \beta_{N-1,N-3}) &= \frac{f_{N-1}'(\hat \beta_{N-1,N-3})g_{N-1}(\hat \beta_{N-1,N-3})-f_{N-1}(\hat \beta_{N-1,N-3})g'_{N-1}(\hat \beta_{N-1,N-3})}{g_{N-1}(\hat \beta_{N-1,N-3})^2},
\end{align*}
to see that 
$$
\tilde T_N'(\hat \beta_{N,N-2}) =\tilde T_{N-1}'(\hat \beta_{N-1,N-3}), \quad N\ge4.
$$
Consequently, $\tilde T_N'(\hat \beta_{N,N-2})$ does not depend on $N$ for $N\ge3$ and it suffices to consider $N=3$. For $N=3$, the common value is explicitly given as 
\begin{align}\label{T3explicitly}
\tilde T_3'( \hat \beta_{3,1}) = -2.07611...
\end{align}

\noindent{\bf Step 4/4:} To prove that $\beta^{(m)}_{N-2}$ defined in \eqref{lineariterationplus} diverges, we iteratively use \eqref{lineariterationplus} to produce 
\begin{align*}
\beta^{(m+1)}_{N-2} -\hat \beta_{N-2} &= \tilde T'( \hat \beta_{N-2})\big(\beta^{(m)}_{N-2} - \hat \beta_{N-2}\big)\\
&= \tilde T'( \hat \beta_{N-2})^2\big(\beta^{(m-1)}_{N-2} - \hat \beta_{N-2}\big)\\
&=... \\
&= \tilde T'( \hat \beta_{N-2})^{m+1}\big(\beta^{(0)}_{N-2} - \hat \beta_{N-2}\big).
\end{align*}
Because $\beta^{(0)}_{N-2} \neq \hat \beta_{N-2}$ and $|\tilde T'( \hat \beta_{N-2})|>1$, we see that 
$\beta^{(m+1)}_{N-2}$ diverges as $m\to\infty$.

2. To see that divergence in coordinate $N-2$ suffices to rule out local stability, we use a first-order Taylor expansion around Kyle's equilibrium $\vec{\hat \beta}\in \R^N$. For $\vec\beta \in \text{dom}(T)$, we have
\begin{align}\label{TaylorN55}
\begin{split}
T(\vec\beta) &= T(\vec{\hat \beta}) + \nabla T(\vec{\hat \beta})(  \vec\beta -\vec{\hat \beta}) + R(  \vec\beta -\vec{\hat \beta}),
\end{split}
\end{align}
where $R$ is the reminder function, which satisfies $R(  \vec\beta ) = o(|  \vec\beta |)$ as $|\vec \beta| \to 0$. Inserting $\vec B$ from \eqref{foverg} into \eqref{TaylorN55} and using \eqref{tildeT} give us
\begin{align}\label{TaylorN5}
\begin{split}
\tilde T(\beta_{N-2}) = \hat \beta_{N-2} + \tilde T'(\hat \beta_{N-2})\big(  \beta_{N-2} -\hat \beta_{N-2}\big) + R(\vec B-\vec{\hat\beta})_{N-2}.
\end{split}
\end{align}

To complete the proof, we argue by contradiction and assume that the sequence $\vec \beta^{(m)}$ defined in \eqref{Kyle_FPm} converges to $\vec{\hat \beta}$. We set $\vec B^{(m)} := (\hat \beta_1,...,\hat \beta_{N-3},\beta^{(m)}_{N-2},\hat \beta_{N-1}, \hat \beta_N )'$. For $m$ sufficiently big,  \eqref{T3explicitly} and $R(\vec\beta ) = o(|  \vec\beta |)$ as $|\vec \beta| \to 0$ give the lower bound
\begin{align}\label{T3explicitly2}
 \bigg|\tilde T'(\hat \beta_{N-2})+\frac{R(\vec B^{(m)} -\vec{\hat\beta})_{N-2}}{(\beta^{(m)} _{N-2} -\hat \beta_{N-2})} \bigg| > 2.
\end{align}
From \eqref{TaylorN5}, we have
\begin{align}\label{TaylorN6}
\begin{split}
\beta^{(m+1)}_{N-2}- \hat \beta_{N-2} &= \tilde T'(\hat \beta_{N-2})\big(  \beta^{(m)}_{N-2} -\hat \beta_{N-2}\big) + R(\vec B^{(m)}-\vec{\hat\beta})_{N-2}\\
&= \bigg(\tilde T'(\hat \beta_{N-2})+\frac{R(\vec B^{(m)}-\vec{\hat\beta})_{N-2}}{(\beta^{(m)}_{N-2} -\hat \beta_{N-2})} \bigg)\big(  \beta^{(m)}_{N-2} -\hat \beta_{N-2}\big).
\end{split}
\end{align}
This gives a contradiction because the iterations \eqref{TaylorN6} diverge by \eqref{T3explicitly2}. 

$\endproof$

\proof[Proof of Theorem \ref{thm3}] Instead of using the $(N-2)$'th coordinate, we redefine $\tilde T$ in \eqref{tildeT} to be the $N$'th coordinate. That is, we define
\begin{align}\label{tildeTN}
\tilde T(\beta^{(m)}_N) :=  T\begin{pmatrix}
\hat \beta_1 \\
\vdots\\
\hat \beta_{N-1} \\
\beta^{(m)}_{N} \\
\end{pmatrix}_{N},
\end{align}
where $(\hat \beta_1,...,\hat\beta_N)$ are Kyle's equilibrium coefficients from Theorem \ref{thm_Kyle} and $\beta^{(m)}_{N}\in \R$ is the variable we iterate in. Proceeding as in the proof of Theorem \ref{lemma_Main}, we see that $\tilde{T}'(\hat \beta_N)=0$ for all $N\ge3$. Consequently, as in the first part of Theorem \ref{lemN2}, the equilibrium value $\hat \beta_N$ is a super-attracting fixed point for $\tilde T$. Proceeding as in the proof of the first part of Theorem \ref{lemN2}, we see that for a starting value $\beta^{(0)}_N$ sufficiently close to $\hat \beta_{N}$ with $\beta^{(0)}_N \neq \hat \beta_N$, the initial policy
$$
\vec B^{(0)} := (\hat \beta_1,...,\hat \beta_{N-1},\beta^{(0)}_N)',
$$
produces a sequence $\vec B^{(m+1)} = T(\vec B^{(m)})=T\big(\cdots T(\vec B^{(0)})\big)$ which converges to $\vec {\hat \beta}$. Because $\vec B^{(0)} \neq \vec{\hat \beta}$, Kyle's equilibrium is not unstable in the sense of Definition \ref{Def_eq3}.

$\endproof$

\end{document}